\documentclass[preprintnumbers,amsmath,amssymb,11pt]{revtex4}
\usepackage{graphicx}
\usepackage{dcolumn}
\usepackage{bm}

\pdfoutput=1
\newcommand\PD[2]{\frac{\partial #1}{\partial #2}}

\newcommand\BO[1]{\mathbf{#1}}

\newcommand{\delsq}{\nabla^2}
\newcommand{\delhsq}{\nabla_h^2}

\newcommand{\ubar}{\overline{u(z)}}
\newcommand{\vbar}{\overline{v(z)}}

\begin{document}

\title{Nonlinear inertia-gravity wave-mode interactions in three dimensional
rotating stratified flows}

\author
{Mark Remmel$^1$, Jai Sukhatme$^{1,2}$ and Leslie M. Smith$^{1,3}$\\
   1. Mathematics Department, University of Wisconsin-Madison, Madison, WI 53706 \\
   2. Indian Institute of Tropical Meteorology, Pashan, Pune 411008, India \\
   3. Engineering Physics Department, University of Wisconsin-Madison, Madison, WI 53706 \\}

\date{\today}

\begin{abstract}

We investigate the nonlinear dynamics of inertia-gravity (IG) wave modes in three-dimensional (3D) rotating 
stratified fluids. Starting from the rotating Boussinesq equations, we derive a reduced partial differential 
equation, the GGG model, consisting of only wave-mode interactions.
We note that this subsystem conserves energy and is not restricted to
resonant wave-mode interactions. In principle,
comparing this model to the full rotating Boussinesq system
allows us to gauge the importance of wave-vortical-wave vs.\ wave-wave-wave interactions in
determining the transfer and distribution of wave-mode energy. As in many atmosphere-ocean
phenomena we work in a skewed aspect ratio domain $H/L$ ($H$ and $L$ are the vertical and horizontal lengths) 
with $Fr = Ro < 1$ such that $Bu = 1$,
where $Fr$, $Ro$ and $Bu$ are the Froude, Rossby and Burger numbers, respectively.
Our focus is on the equilibration of wave-mode energy
and its spectral scaling under the influence
of random large-scale ($k_f$) forcing. We present results from two sets of parameters:
(i) $Fr=Ro \approx 0.05$, $H/L=1/5$, and
(ii) $Fr=Ro \approx 0.1$, $H/L=1/3$. As anticipated from prior work, when forcing is applied
to all modes with equal weight, with
$Fr=Ro \approx 0.05$ and $H/L=1/5$, the wave-mode energy
of the full system equilibrates and its spectrum scales as a power-law that lies between 
$k^{-1}$ and $k^{-5/3}$ for $k_f < k < k_d$, where $k_d$ is the dissipation scale. 
For the same parameters, when forcing is restricted to only wave modes, the 
wave-mode energy fails to equilibrate in both the full system as well as the GGG subsystem
at the resolutions we can achieve. 
This cleary demonstrates the importance of the vortical mode (by facilitating wave-vortical-wave interactions)
in determining the wave-mode
energy in the rotating Boussinesq system.
Proceeding to the second set of simulations, i.e.\ for the larger $Fr=Ro\approx 0.1$
in a less skewed aspect ratio domain with $H/L=1/3$, we observe that the energy of the
GGG subsystem equilibrates and is resolution independent. Further, the full system with forcing restricted to 
wave modes also equilbrates and both yield identical energy spectra.
Thus it is clear that
the wave-wave-wave interactions play a role in the overall dynamics at
moderate $Ro$, $Fr$ and aspect ratios. 
Apart from theoretical concerns addressed in the Conclusion, from a practical standpoint these results 
highlight the difficulty in 
properly resolving wave-mode 
interactions when simulating realistic geophysical phenomena.

\end{abstract}

\maketitle

\section{Introduction}

Inertia gravity (IG) waves, resulting from the rotating and stratified nature of geophysical fluids, 
play an important role in atmosphere-ocean dynamics \cite{Gill}.  A broad overview of their properties
with relevance to the ocean can be found in Garrett \& Munk \cite{Garrett-Munk}, and a recent 
review focussing on the observational characterization of the oceanic wave-field can be found
in Polzin and Lvov \cite{PL}.  The general dynamics of these waves are reviewed by Sommeria \& Staquet \cite{S-rev}.  Further, Wunsch \& Ferrari \cite{Wunsch-Ferrari} place these waves in context when considering the general circulation of the ocean, while Fritts \& Alexander \cite{Fritts-Alexander} review their influence on diverse phenomena in the middle atmosphere. \\

In their influential papers, Garrett \& Munk 
\cite{Garrett-Munk},\cite{GM1} pointed out 
the importance of a deeper understanding of the dynamics of the IG wave modes.  
In this regard, one avenue of progress has been to focus upon the subset of resonant interactions 
among these waves \cite{Mc-B}.      In general, the dispersive nature of the linear
waves results in reduced nonlinear transfer between modes by dispersive
phase scrambling.  However, resonant interactions are special nonlinear
interactions for which phase scrambling is absent, and nonlinear transfer remains
strong.  Thus resonant interactions are critical for the dynamics when they are present.
See for example \cite{Anile} for a broad introduction to nonlinear 
dispersive equations, \cite{Zak} for statistical theories of the so-called weak turbulence, \cite{Majda-book}, 
\cite{Chemin-book} 
for geophysical perspectives and \cite{Bartello},\cite{SS-gafd} for the
roles of resonances and near-resonances in the specific case of the 3D rotating Boussinesq system. \\

Investigating resonant IG waves, McComas \& Bretherton \cite{Mc-B} distinguished
between three classes of resonant interactions that lead to induced diffusion, 
elastic scattering and parametric 
sub-harmonic instability. The picture put forth was, 
for a large-scale initial data:
(i) elastic scattering 
leads to 3D isotropy of an initial condition that is only horizontally isotropic,
(ii) sub-harmonic instability moves energy downscale, (iii) induced diffusion
tends to force the high wavenumber spectrum towards equilibrium.  A detailed account of these 
resonances can be found in the extensive review by Muller 
et al.\ \cite{Muller-rev}; their review also describes approaches 
that account for more than resonant interactions, but involve other 
approximations (such as the direct interaction approximation).  
Further, the use of resonances to derive kinetic equations under the weak
turbulence paradigm has been an active area of work --- for an overview see Lvov et\ al.\ \cite{L-web}.
In particular recent work has 
shown a correspondence between observed IG energy spectra and 
particular steady state solutions of the 
relevant kinetic equations \cite{L1},\cite{L2}. \\

Unfortunately, as was pointed out by McComas \& Bretherton \cite{Mc-B}, resonances 
do not account for the bulk of interactions among IG waves.  A formal 
statement regarding the sparsity of these resonances can be found in 
Babin et\ al.\ \cite{Babin-rev}.  This is all the more 
an important issue when $Ro,Fr$ are small (implying strong rotation and stratification) but 
far from zero, i.e.\ $0\ll Ro,Fr < 1$ (where $Ro,Fr$ are the Rossby and Froude numbers respectively). 
Indeed, having non-zero $Fr$ and/or $Ro$ opens the door for near-resonances to enter the dynamics 
and these interactions can, both quantitatively and qualitatively, change the
behavior of the system \cite{SS-gafd}.  This naturally motivates the construction of a model 
(called the GGG model) that includes all possible wave-mode interactions.  Quite interestingly, as the two-dimensional (2D) stratified Boussinesq system only supports 
wave modes, this new 3D system can be looked upon as an extension of 
the full 2D stratified problem
\cite{BS},\cite{SS},\cite{Smith-short}. \\

So far, our discussion has been restricted to considering 
wave modes in isolation.  In reality
rotating and stratified fluids support an additional vortical mode of motion \cite{LR}, and in fact
in geophysically relevant limits, interactions among vortical modes lead to the 
celebrated quasigeostrophic (QG) equations 
\cite{Bartello},\cite{Babin-rev},\cite{SW}.  Indeed, one of the many contributions by Prof.\ Majda (in collaboration with Prof.\ P.\ Embid) 
in the field of geophysical fluid dynamics concerns a formal statement on the emergence of QG dynamics, from the 
governing rotating Boussinesq equations, in the limit
$Ro \sim Fr = \epsilon \rightarrow 0$ while holding $Bu \sim 1$ \cite{Embid1998},\cite{Babin-rev}.  In addition to clarifying the 
nature of the QG equations and constructing a general framework for averaging over fast-modes in 
geophysical systems \cite{Embid1996}, their
work also predicted the emergence of a new regime, the so-called 
vertically sheared horizontal flows (VSHF) when $Ro \sim 1$ while $Fr = \epsilon \rightarrow 0$ 
\cite{Embid1998},\cite{MajdaEmbid}.  Since then, numerical work in the appropriate parameter regime has confirmed
the emergence of VSHF modes when 
considering a rotating Boussinesq fluid under random forcing \cite{SW},\cite{Laval},\cite{WB1},\cite{WB2}. \\

As it happens, the vortical mode plays an important role in the redistribution of
wave-mode energy by means of the so-called wave-vortical-wave 
interactions. In fact, this is thought to be the primary manner in which energy is transferred from 
large to small scales in rotating and stratified flows, and is attributed to be the fundamental 
mechanism behind geostrophic adjustment \cite{Bartello}. Further, in both 
decaying and forced scenarios, numerical 
simulations of rapidly rotating and strongly stratified (i.e.\ small $Ro,Fr$) flows have shown 
the equilibration 
of wave modes (by means of the aforementioned forward transfer)
while energy continues to be transferred 
upscale in the vortical modes via vortical-vortical-vortical (QG) interactions \cite{Bartello},\cite{SS-gafd}.
For $f=N$ (where $f,N$ are the Coriolis parameter and the Brunt-V\"ais\"al\"a frequency) in a unit
aspect ratio, the 
wave modes are passively driven by the vortical modes.  For this case $f=N$, 
there are no wave-wave-wave 
resonances (or near-resonances), and 
the distribution of energy among the wave modes is 
strongly influenced by the presence of the vortical mode \cite{Bartello},\cite{SS-gafd}. 
Of course, in the atmosphere-ocean system $f \neq N$, 
leading to the possibility of near and exact wave-wave-wave resonances, though as 
stated earlier these sets of interactions are sparse \cite{Babin-rev}. 
All in all, this line of reasoning leads one to believe that the vortical mode should in fact
play a significant role in determining the wave-mode energy distribution. \\

These scenarios then lead to a somewhat dichotomous state of affairs with regard to IG wave-mode energy, i.e.\
on one hand we have theories based on the interactions of IG modes in isolation.
On the 
other hand, there is reason to attribute a prominent role to the vortical mode in the overall
dynamics of a rotating and stratified fluid, which includes the distribution of energy among the 
wave modes themselves. This dichotomy --- in the context of purely stratified flows --- 
has been pointed out and investigated by Waite \& Bartello 
\cite{WB2}.   By forcing only wave modes of the Boussinesq system, they inquired into the 
turbulence generated by
a stratified fluid. Their principal conclusion was to highlight the importance of the vortical mode in
mediating wave-mode energy transfer and distribution. Further,
they noted the difficulty in properly resolving wave-mode interactions \cite{WB2}. 
In the present work, in addition to studying the full Boussinesq equations, we derive a model that consists of 
only wave-mode interactions,  
which in principle provides the means for determining the relative importance of wave-vortical-wave
and wave-wave-wave interactions for the distribution of wave-mode energy. 
Specifically, by simulating the newly derived GGG model, we attempt to see if this system 
is capable of a robust forward transfer of energy, and compare the energy spectra of wave modes from the
GGG model to the corresponding wave mode spectra that arise from simulations of the full rotating-stratified 
Boussinesq system.
We proceed to introduce the basic equations, review the possible means of wave mode evolution and derive the 
GGG model. We then describe the numerical setup and proceed to the simulations of this system. 
Finally, we collect 
and discuss the results. 

\section{Basic Equations}

The rotating Boussinesq equations in a 3D periodic setting are \cite{Majda-book}

\begin{eqnarray}
\frac{{\rm D} {\bf u}}{{\rm D}t} + f \hat{z} \times {\bf u} = - \nabla \varphi - N \theta \hat{z} + \nu \nabla^2 {\bf u} \nonumber \\
\frac{{\rm D} \theta }{{\rm D}t} - N w = \kappa \nabla^2 \theta ~,~ \nabla \cdot \BO{u}=0
\label{1}
\end{eqnarray}
where ${\bf u}=(u,v,w)$ is the 3D velocity field, $\varphi=p/\rho_0$ is a scaled pressure and $f$ is the 
Coriolis parameter
with rotation assumed to be aligned with the $z$-axis.  Equations (\ref{1}) result from
considering periodic perturbations to a state of hydrostatic balance wherein the density profile satisfies
$\rho = \rho_0 - b z + \rho'$ with $|\rho'|, |bz| \ll \rho_o$.
Further, we have set $\rho'= (N \rho_o/g) \theta$ where $N=(gb/\rho_0)^{1/2}$.
Finally, $\nu$ and $\kappa$ are the viscosity and diffusivity respectively. \\

Linearizing (\ref{1}) about a state of rest, substituting solutions of the form $\phi = \hat{\phi} \exp\{{\rm i} {\bf k}\cdot{\bf x}
-\sigma({\bf k}) t \}$ one finds

\begin{equation}
\sigma_{\pm} ({\bf k}) = \pm  \frac{ ( N^2 k_h^2 + f^2 k_z^2)^{1/2}}{k}~;~ \sigma_0({\bf k})=0.
\label{2}
\end{equation}
Here $k$ is the magnitude of the wave-vector and $k_h^2=k_x^2+k_y^2$. 
Modes corresponding to $\sigma_{\pm}$ are the IG waves (referred to
as wave modes), 
while the mode corresponding to $\sigma_0$ is referred to as
the vortical (or geostrophic) mode \cite{LR},\cite{Bartello},\cite{SW}.
Denoting the eigenfunctions corresponding to $\sigma_0,\sigma_{\pm}$ by $\phi_0,\phi_{\pm}$ respectively, 
it is known that these eigenfunctions are mutually orthogonal and form a complete basis 
(see Embid \& Majda \cite{Embid1996},
\cite{Embid1998} and Smith \& Waleffe \cite{SW} for details). This enables us to project a
Fourier expansion of the solution to (\ref{1}) onto this eigenfunction basis, i.e.\ projections of the form
\begin{equation} 
\BO{v}({\bf k},t) = a_0({\bf k},t) \phi_o({\bf k}) + a_+({\bf k},t) \phi_+({\bf k}) + a_-({\bf k},t) \phi_-({\bf k}),
\label{represent}
\end{equation} 
where $\BO{v}$ is the state vector, $\BO{v}^T=(u,v,w,\theta)$ \cite{Embid1998}, \cite{Bartello}, \cite{SW}. Symbolically this yields 

\begin{equation}
\PD{a_\alpha(\BO{k},t)}{t}+{\rm i}\sigma_\alpha(\BO{k}) a_\alpha(\BO{k},t)=\sum_\triangle\sum_{\beta,\gamma}C_{\BO{k}\BO{p}\BO{q}}^{\alpha\beta\gamma}a_\beta(\BO{p},t) a_\gamma(\BO{q},t)
\label{3}
\end{equation}
where $\triangle$ represents a sum over ${\bf p},{\bf q}$ such that ${\bf k}={\bf p}+{\bf q}$. The indices $\alpha,\beta,\gamma$ run over $0,+,-$ and $C_{{\bf k}{\bf p}{\bf q}}^{\alpha \beta \gamma}$ is the interaction coefficient.  The exact form of the interaction coefficients is not emphasized here, but it is helpful in our discussion to think in terms of the following symmetric definition
\begin{equation}
C_{{\bf k}{\bf p}{\bf q}}^{\alpha \beta \gamma}=\frac{-{\rm i}}{2}\left((\phi_{\beta}^{\BO{u}}(\BO{p})\cdot\BO{q})\phi_{\gamma}(\BO{q})+(\phi_{\gamma}^{\BO{u}}(\BO{q})\cdot\BO{p})\phi_{\beta}(\BO{p})\right)\cdot\phi_{\alpha}^{*}(\BO{k}),
\label{intcoeff}
\end{equation}
where $\phi_{\beta}^{\BO{u}}$ contains the first three components of $\phi_{\beta}$ corresponding to the velocity entries and $(\cdot)^{*}$ denotes complex-conjugate.  Notice 
that, with this symmetric definition, $C_{{\bf k}{\bf p}{\bf q}}^{\alpha \beta \gamma}=C_{{\bf k}{\bf q}{\bf p}}^{\alpha \gamma \beta}$; the
coefficient $C_{{\bf k}{\bf p}{\bf q}}^{\alpha \beta \gamma}$ involves the advection of the mode 
corresponding to $q$ by the mode corresponding to $p$ and vice versa.  In the following, a triplet $(\cdot, \cdot, \cdot)$ refers to a particular type of triad interaction, and a
$v$ refers to a vortical mode while a $w$ refers to a wave mode (either a $+$ or a $-$ mode). 
For example, $(w,w,v)$ (with all permutations implied) represents all possible interactions between two wave modes
and one vortical mode.

\subsection{Wave-mode evolution}

An alternate way to represent the expansion is 
$\BO{v}({\bf k},t)=b_0({\bf k},t)\phi_o({\bf k})\exp[-{\rm i}\sigma_0(\BO{k})t]+
b_+({\bf k},t)\phi_+({\bf k})\exp[-{\rm i}\sigma_+(\BO{k})t]+
b_-({\bf k},t)\phi_-({\bf k})\exp[-{\rm i}\sigma_-(\BO{k})t]$. In this case (\ref{3}) takes the form

\begin{equation}
\PD{b_\alpha(\BO{k},t)}{t}=\sum_\triangle\sum_{\beta,\gamma}C_{\BO{k}\BO{p}\BO{q}}^{\alpha\beta\gamma}b_\beta(\BO{p},t) b_\gamma(\BO{q},t)\exp\{{\rm i}[\sigma({\bf k})-\sigma({\bf p})-\sigma({\bf q})]~t\}
\label{3a}
\end{equation}
which shows the influence of the dispersion on the advective nonlinearity. 
Indeed, it is now possible to construct a dynamical hierarchy consisting of (i) resonances, i.e.\
$|\sigma({\bf k})-\sigma({\bf p})-\sigma({\bf q})|=0$, (ii) near-resonances, 
i.e.\ $0 < |\sigma({\bf k})-\sigma({\bf p})-\sigma({\bf q})| \le \delta$ where $\delta=\min{(Ro,Fr)}$ and
(iii) non-resonant interactions, i.e.\ $|\sigma({\bf k})-\sigma({\bf p})-\sigma({\bf q})|>\delta$.
In this hierarchy it is postulated that resonances are always important, near-resonances play 
a role when $0 < \delta < 1$ i.e.\ rotation and stratification are strong, but we are outside
limiting scenarios where $\delta \rightarrow 0$. Finally, for rapid rotation and strong stratification,
non-resonant interactions are only expected to contribute at higher-orders (in 
$\delta$). While it is not possible for non-resonant interactions to lead to qualitatively different behavior as 
compared to 
near-resonances, we note that near-resonances open up pathways that are precluded under 
resonance (for details see Sukhatme \& Smith \cite{SS-gafd}).
An explicit example of such a qualitative change induced by near-resonant interactions is provided via cyclonic-anticyclonic 
asymmetry in rotating flows \cite{Smith-Lee}.  Other examples, such as the influence on the statistical features of 
water waves and capillary wave generation may be found in Annenkov \& Shrira \cite{AnSh}, 
and Watson \& Buschbaum \cite{WaBu}, respectively. \\

Considering only resonant and near-resonant interactions, 
the wave modes can only evolve via the $(w,w,v)$ and $(w,w,w)$ classes
(see Bartello \cite{Bartello} 
or Sukhatme \& Smith \cite{SS-gafd} for a detailed view of wave and vortical mode evolution). 
When $1/2 \le f/N \le 2$ the $(w,w,w)$ (near) resonant interactions are absent and --- within 
this dynamical hierarchy --- the 
distribution of wave-mode energy is determined by $(w,w,v)$, indicating the crucial role played by the
vortical mode. Further, it was shown by Bartello \cite{Bartello}, that purely resonant $(w,w,v)$ 
interactions are catalytic in that they involve
the exchange of energy among two wave modes by using the vortical mode as a catalyst. This 
leads to the elegant result that when $f=N$ the wave modes are passively driven by the vortical mode
\cite{Bartello}
(this follows as near-resonances are not possible for a rapidly rotating and strongly stratified 
fluid when $f=N$ \cite{SS-gafd}). 
Of course, in the most general case we lift the restriction $1/2 \le f/N \le 2$. As both 
$(w,w,v)$ and $(w,w,w)$ are viable means of affecting the wave-mode energy distribution, 
it is natural to inquire into their nature and importance. Specifically, given the formal sparsity of 
$(w,w,w)$ interactions \cite{Babin-rev}, are $(w,w,v)$ always dominant? Indeed, do the 
$(w,w,w)$ interactions play any role in the 3D rotating Boussinesq system? Also, what is 
the nature of these interactions, i.e.\ do they yield a forward transfer of energy under 
large-scale forcing? If so, do they lead to an equilibration of wave modes and yield 
a well-defined power spectrum? To inquire into these questions, we construct a model (the 
GGG model) that includes only wave-mode interactions. In principle, comparing the 
results from the GGG model with the full rotating Boussinesq system should help in settling some 
of these issues.

\subsection{The GGG model}

We now present the GGG model in physical space and its derivation.  It is useful to first determine the mode amplitudes in terms of the physical variables.  Towards this goal, begin by defining the stream function $\psi$, the potential function $\chi$, and the horizontally averaged flows $\overline{u(z)}$ and $\overline{v(z)}$ in the standard way

\begin{equation}
u=\chi_x-\psi_y+\overline{u(z)}\quad\text{and}\quad v=\chi_y+\psi_y+\overline{v(z)}.
\label{alternate}
\end{equation}
Here $\overline{(\cdot)}$ denotes a horizontal average. These definitions imply $\nabla_h^2\chi=u_x+v_y$ and $\nabla_h^2\psi=v_x-u_y$. 
Substituting (\ref{alternate}) for $u$ and $v$ in the state vector ,$\BO{v}$, (\ref{represent}) and the orthonormality of 
the eigenfunctions yields (please see the Appendix for additonal details)

\begin{eqnarray}
\label{azero1}
a_k^0=\frac{{\rm i}N}{\sigma_kk} (-k_h^2\psi- {\rm i}\frac{f}{N}k_z\theta) \\
\label{aplus1}
a_k^+=\frac{1}{\sqrt{2}\sigma_kk}(-\frac{\sigma_k w_k k^2}{k_h}-fk_z k_h\psi+ {\rm i}Nk_h\theta) \\
\label{aminus1}
a_k^-=\frac{1}{\sqrt{2}\sigma_kk}(-\frac{\sigma_k w_k k^2}{k_h}+fk_z k_h\psi- {\rm i}Nk_h\theta) \\%
\label{vshfamps1}
a_{k_z}^0=\overline{\theta(z)}\text{sgn}(fk_z)\quad a_{k_z}^+=\frac{-{\rm i}\overline{u}+\overline{v}}{\sqrt{2}}
\quad a_{k_z}^-=\frac{{\rm i}\overline{u}+\overline{v}}{\sqrt{2}}.
\end{eqnarray}
Next, notice that by adding $a_k^+$ and $a_k^-$ and also subtracting $a_k^-$ from $a_k^+$ the physical variables decouple further, 

\begin{eqnarray}
\label{apluspaminus1}
a_k^++a_k^-=-\frac{\sqrt{2}kw_k}{k_h}\\
\label{aplusmaminus1}
a_k^+-a_k^-=\frac{\sqrt{2}{\rm i}fk_h}{\sigma_kk}(\frac{N}{f}\theta+{\rm i}k_z\psi)\\
\label{modevshf1}
(a^+_{k_z}+a^-_{k_z})=\sqrt{2}\overline{v(z)}\qquad (a^+_{k_z}-a^-_{k_z})=-{\rm i}\sqrt{2}\overline{u(z)}.
\end{eqnarray}
In fact, it is now clear from (\ref{azero1}) and (\ref{aplusmaminus1}) that defining

 \begin{equation}
M=\nabla_h^2\psi-\frac{f}{N}\PD{\theta}{z} \quad\text{and}\quad R=\frac{N}{f}\theta+\PD{\psi}{z},\\
\label{g1}
\end{equation}
is advantageous since this means 
 \begin{equation}
a_k^0=\frac{iN}{\sigma_kk}M_k \quad\text{and}\quad a_k^++a_k^-=\frac{\sqrt{2}ifk_h}{\sigma_kk}R_k.
\label{associate}
\end{equation}
In addition $M$ and $R$ have physical significance : $M$ is the linear potential vorticity (PV) and $R$ is a measure of the 
geostrophic imbalance. It is apparent that the vortical modes contain all the linear PV and that the variable $R$ is 
associated with the IG modes \cite{Bartello}. In the following it will prove useful to make the following definitions

\begin{equation}
\O=(\nabla^2_{h}+\frac{f^2}{N^2}\partial_{zz}) ~~;~ A=\O^{-1}M,~S=\O^{-1}R.
\label{notation}
\end{equation}
Using these definitions note that (\ref{g1}) can be re-arranged in the form

\begin{equation}
\psi=A+\frac{f^2}{N^2}S_z ~,~ \theta=\frac{f}{N}\nabla^2_hS-\frac{f}{N}A_z.
\label{g2}
\end{equation}
A systematic method to complete the derivation of the GGG model in physical space involves the three equations obtained from 
(\ref{3}), one for each $\alpha$ ($0,+,-$), after retaining only the interactions among the IG modes. 
Therefore, when $\alpha$ is 0 there are no interactions and the right hand side is zero; whereas, when $\alpha$ is $+$ or $-$, 
the right hand side is modified by restricting the sum involving $\beta$ and $\gamma$ to run over $+$ and $-$ only. 
Then in order to substitute in the physical space variables identified with $a_k^+\pm a_k^-$ (above), add and subtract the 
equations corresponding to $\alpha$ being $+$ and $-$. Finally substitute the physical variables for the mode amplitudes and 
inverse Fourrier transform the equations. This process involves a fair amount of algebra 
and relevant details are presented in the Appendix. \\

Here, for the sake of brevity, we present a more illuminating and much shorter derivation. This is  possible now that the 
appropriate physical space variables corresponding to the vortical modes as well as the IG modes have been identified. 
It is convenient to rewrite the original system given by (\ref{1}) in a form that has a separate equation for the time 
derivative of each of the variables corresponding to the modal amplitudes. A consistent system of this sort contains a time 
derivative for 
$M$ corresponding to the vortical modes (\ref{azero1}), and time derivatives for $f\nabla_h^2R$, $\nabla^2w$, $\overline{u(z)}$, 
and $\overline{v(z)}$ which all correspond to the IG modes (see (\ref{apluspaminus1},(\ref{aplusmaminus1}) and (\ref{modevshf1})). 
This reformulation of (\ref{1}) results in

\begin{eqnarray}
\PD{M}{t}+\hat{z}\cdot\nabla\times(\BO{u}\cdot\nabla\BO{u})-
\frac{f}{N}\partial_z[(\BO{u}\cdot\nabla)\theta]=0, \nonumber \\
\PD{f\delhsq R}{t}-N^2\O w +N\delhsq[(\BO{u}\cdot\nabla)\theta]+
f\partial_z(\hat{z}\cdot\nabla\times(\BO{u}\cdot\nabla\BO{u}))=0, \nonumber \\
\PD{\delsq w}{t}+f\delhsq R+\delhsq(\BO{u}\cdot\nabla w)-
\partial_z(\nabla_h\cdot(\BO{u}\cdot\nabla\BO{u}_h))=0, \nonumber \\
\PD{\ubar}{t}-f\vbar+\overline{\partial_z(uw)}=0, \nonumber \\
\PD{\vbar}{t}+f\ubar+\overline{\partial_z(vw)}=0.
\label{g3}
\end{eqnarray}
Here, the $M$ equation in (\ref{g3}) is obtained by subtracting $f/N$ times the $z$ derivative of the $\theta$ equation,

\begin{equation}
\frac{\partial\theta_z}{\partial t}+\partial_z((\mathbf{u}\cdot\nabla)\theta)-Nw_z=0,
\label{zdtheta}
\end{equation}
from the vertical component of vorticity equation,

\begin{equation}
\frac{\partial\nabla_h^2\psi}{\partial t}+\hat{z}\cdot\nabla\times(\BO{u}\cdot\nabla\BO{u})-fw_z=0.
\label{vertvort}
\end{equation}
The $R$ equation follows by adding $N$ times the the horizontal Laplacian of the $\theta$ equation,

\begin{equation}
\PD{\nabla^2_h\theta}{t}+\nabla_h^2((\mathbf{u}\cdot\nabla)\theta)-N\nabla_h^2w=0,
\label{hltheta}
\end{equation}
to $f$ times the $z$-derivative of the vertical vorticity equation (\ref{vertvort}).
The equation for the Laplacian of $w$ in (\ref{g3}) involves first solving for $\varphi$ in (\ref{1}) by taking the divergence 
of the momentum equations and applying the incompressibility constraint, this yields the formal relation

\begin{equation}
\varphi=\nabla^{-2}(-N\theta_z-\nabla\cdot(\BO{u}\cdot\nabla\BO{u})+f\nabla_h^2\psi).
\label{pressure}
\end{equation}
Now, taking the Laplacian of the $w$ equation in (\ref{1}) using (\ref{pressure}) for $\phi$ gives the relevant equation in 
(\ref{g3}). 
Note that in taking the required derivatives we lost information about the time derivatives of horizontally averaged 
$u$ and $v$ fields; therefore, we must include the horizontally averaged time derivative equations for $u$ and $v$ in (\ref{1}) 
to complete the reformulation (\ref{g3}). 
Note that the horizontally averaged information for $\theta$ is contained in the $M$ equation; the vortical modes for 
$k_h=0$ do not need separate consideration \footnote{Technically domain averaged equations for $w$ and $\theta$ should be added 
to the reformulation in (\ref{g3}) . 
However, although the system in (\ref{1}) is not Galilean invariant, a domain average of the $u$, $v$, $w$, and $\theta$ 
equations of (\ref{1}) shows that mean flows will be absent for all time if not present initially nor forced.}. \\
 
It is clear that the GGG model, i.e.\ a dynamical system that only involves wave-mode interactions should not admit any 
interactions involving $M$.  This is most easily achieved by using new variables 
$(u',v',w',\theta')$ where

\begin{eqnarray}
u' \equiv\chi_x-\frac{f^2}{N^2}S_{zy}+\overline{u(z)} ~,~ v'\equiv\chi_y+\frac{f^2}{N^2}S_{zx}+\overline{v(z)}
\nonumber \\
w'\equiv w ~,~ \theta' \equiv\frac{f}{N}\nabla_h^2 S
\label{g4}
\end{eqnarray}
On comparing the primed variables in (\ref{g4}) to the original ones in (\ref{g3}) we see that the new variables achieve the goal 
of removing terms involving the vortical mode from the advective nonlinearity. Further, since all interactions involving 
a vortical mode are to be removed, the evolution of $M$ itself must be set to zero. This results in the following concise 
form for the GGG model 

\begin{eqnarray}
\PD{M}{t}=0 \nonumber \\
\PD{f\delhsq R}{t}-N^2\O w +N\delhsq[(\BO{u'}\cdot\nabla)\theta']+
f\partial_z(\hat{z}\cdot\nabla\times(\BO{u'}\cdot\nabla\BO{u'})=0 \nonumber \\
\PD{\delsq w}{t}+f\delhsq R+\delhsq(\BO{u'}\cdot\nabla w)-
\partial_z(\nabla_h\cdot(\BO{u'}\cdot\nabla\BO{u'}_h))=0 \nonumber \\
\PD{\ubar}{t}-f\vbar+\overline{\partial_z(u'w)}=0 \nonumber \\
\PD{\vbar}{t}+f\ubar+\overline{\partial_z(v'w)}=0
\label{g5}
\end{eqnarray}
Note that (\ref{g5}) are well-defined PDEs in physical space, in fact they also conserve energy \cite{Kr} 
(see Remmel \& Smith \cite{RS} for a detailed discussion).  

\section{Numerical Results \& Discussion }

The Rotating Boussinesq Equations and the GGG model, i.e.\ (\ref{g3}) and (\ref{g5}) respectively, are solved using pseudo-spectral codes in a triply periodic box of dimensions $H\times L \times L$, implying $H$ ($L$) is the vertical height (horizontal length) of the domain.  The time stepping is done using a third order Runge-Kutta scheme.  Energy is removed at small scales via an eighth-order hyperviscosity term, and at large scales via a linear relaxation term.  The combination has the following generic form in Fourier Space
\begin{equation}
\PD{\hat{y}(\BO{k})}{t}+ .....=(-\nu k^{16}-\alpha)\hat{y}(\BO{k}),
\end{equation}
with $\alpha=0.1*I\{k\le2.5\}$ and
\begin{equation}
\nu=2.5\left(\frac{E(k_m,t)}{k_m}\right)^{1/2}k_m^{2-16}.
\label{nu}
\end{equation}
In (\ref{nu}) $k_m$ is the highest available wavenumber and $E(k_m,t)$ is the value of the 
energy spectrum at wavenumber $k_m$.  All linear terms, including the aforementioned diffusion terms, were treated with an integrating factor.  \\

The numerical results were obtained by forcing each system from a state of rest.  The forcing is uncorrelated in time, with
wavenumber spectrum $F(k)$ given by a Gaussian with standard deviation $s=0.5$ and energy input rate $\epsilon=1$, 
\begin{equation}
F(k)=\epsilon\frac{\text{exp}(-0.5(k-k_f)^2/s^2)}{\sqrt{2\pi}s}.
\label{force}
\end{equation} 
The peak wavenumber $k_f$ is chosen as $k_f=4$ for all results reported herein.   
The non-dimensional parameters are defined as $Fr=U/(NL_z)$ and $Ro=U/(fL_h)$, where $U$ is derived from the energy input rate and the scale at which the energy enters the system, i.e.\ $U=(\epsilon/k_f)^{1/3}$. The length scales $(L_z,L_h)$ are consistent with $U$, i.e.\
$L_z=1/k_f$ and $L_h$ follows from the prescribed aspect ratio $L_h=L*L_z/H$. 
As in atmosphere-ocean phenomena, we set $H/L < 1$ and $f<N$.  In fact, 
we choose $f,N$ to offset the skewed aspect ratio ($H/L$) so as to satisfy $Fr=Ro \Rightarrow Bu = 
Ro^2/Fr^2 = 1$. The numerical resolution of the simulations is chosen to yield an isotropic grid, for 
example if $H/L=1/5$, we use five times as many points in the horizontal directions. \\

The modal decomposition in (\ref{represent}) is used to feed energy into the system by directly forcing the mode amplitudes.  The full system is forced by one of the following two methods :  

\begin{enumerate}
\item  All three mode amplitudes $a_0$, $a_\pm$ are forced with equal weight (see (\ref{3})).
\item  The amplitude $a_0$ is set to zero at every time step such that only wave modes are excited.
\end{enumerate}

The latter forcing is always used 
with the GGG model since there are no interactions involving vortical modes.
Note that for fixed $\epsilon=1$ in (\ref{force}), the computed energy input rate for method 2 will be reduced by a
factor 2/3 as compared to method 1,  changing the $Ro$ and $Fr$ numbers
 by a factor of $(2/3)^{1/3}=0.87$.   We present the comparison of GGG and full model simulations 
 {\it keeping the same 
 level of forcing for the wave-mode amplitudes}, rather than the same $Ro$ and $Fr$ numbers.  
 Thus, apart from the actual random numbers, the forcing of the IG modes is always identical for the
 two methods and for the two systems (full and GGG).   The results will be interpreted keeping
 this choice in  mind.\\

\begin{figure}
\centering
\includegraphics[width=10cm,height=10cm]{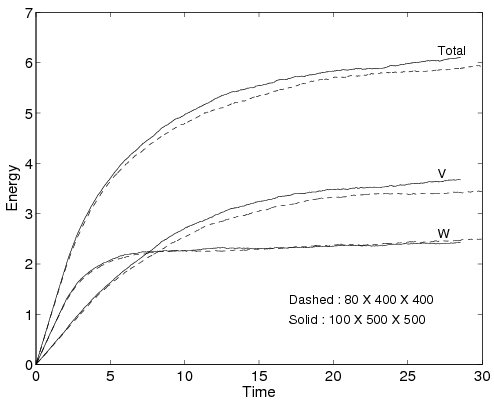}
\caption{\label{fig1} The wave-mode, vortical-mode and total energy vs. time for the full system. The 
simulations are at resolutions of $80 \times 400 \times 400$ and $100 \times 500 \times 500$ in an aspect
ratio of $1/5$ and $Ro=Fr=0.05$. The runs are
halted when the vortical modes near equilibration. }
\end{figure}

\begin{figure}
\centering
\includegraphics[width=10cm,height=10cm]{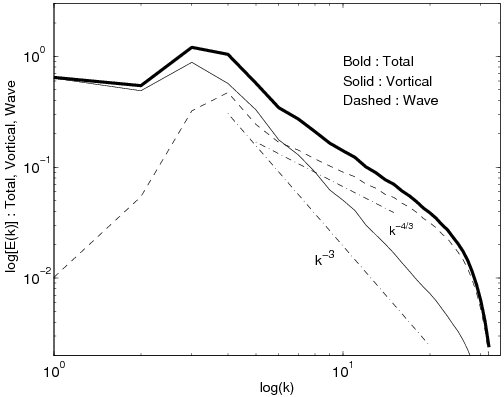}
\caption{\label{fig2} The wave-mode, vortical-mode and total energy spectrum (at $t \approx 28$). The
results are from the $100 \times 500 \times 500$ run. Note that the scaling of the vortical modes 
is consistent with a enstrophy-cascading $k^{-3}$ form for $k_f < k < k_d$. The wave-mode spectrum
is best described by a $k^{-4/3}$ scaling --- we do not attribute any special significance to this number. 
Rather, it is drawn to show that the wave-mode scaling is consistent with a power law between $-1$ and $-5/3$.}
\end{figure}

\begin{figure}
\centering
\includegraphics[width=8cm,height=8cm]{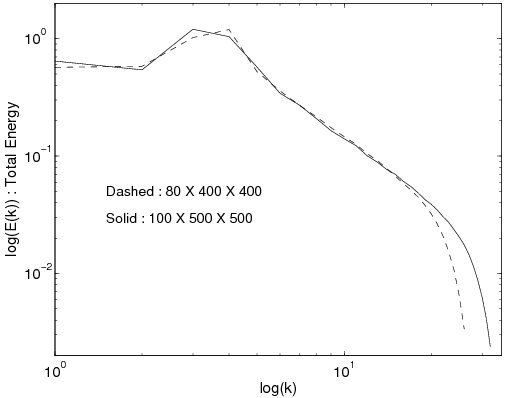}
\includegraphics[width=8cm,height=8cm]{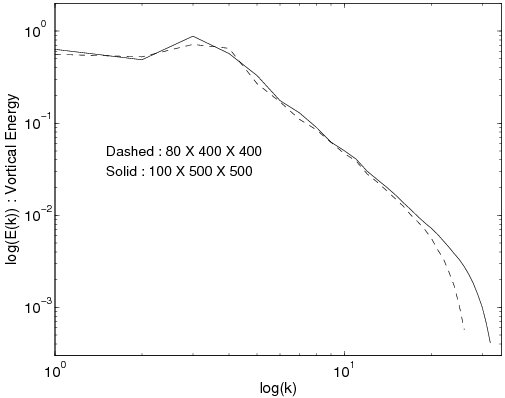}
\includegraphics[width=8cm,height=8cm]{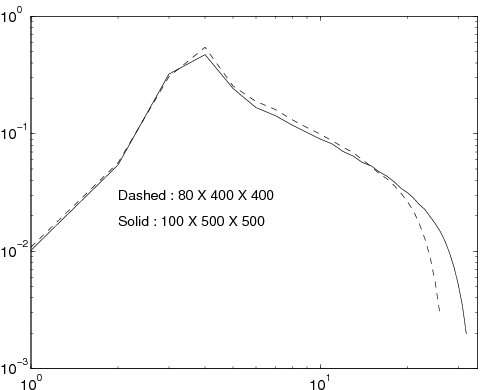}
\caption{\label{fig3} A comparison of the spectra for the full system at the two different resolutions.
The simulations appear to have converged and are insensitive to the increasing resolution.}
\end{figure}

Given our present computational resources, $H/L=1/5$ is the most skewed aspect ratio we studied (estimates for mid-latitude atmosphere-ocean dynamics lead to $1/10 \le f/N \approx H/L \le 1/100$,
but unfortunately these small ratios are well beyond our computational capabilities).  Results obtained using $H/L=1/5$ are presented first.  The full system, forced using method 1, is studied at 
resolutions of $80 \times 400 \times 400$ and $100 \times 500 \times 500$ under rapid rotation and strong stratification (specifically, $Ro=Fr=0.05$).  As is evident in Fig.\ (\ref{fig1}), the wave-mode energy saturates after about 5 (dimensional) time units, and is then almost flat for the next 25 time 
units.  In this 
parameter regime, one expects the energy to be transferred upscale among the vortical modes, however the large-scale damping allows these modes to equilibrate (see \cite{Babin-rev}, \cite{Bartello}, 
\cite{SS-gafd} for analytical and numerical 
work on the nature of the various mode interactions when $Bu \sim O(1)$). 
We choose to halt the simulations as soon as the 
vortical modes near equilibration, because from then onwards the damping directly affects the dynamics. 
Similarly,
if one does not include large-scale damping, then the build-up of energy in the vortical modes soon 
leads to finite-size effects.
 In essence,
one has a small temporal window (depending on the forcing scale and the $Fr,Ro$ numbers) 
during which to study the characteristics of the different modes interactions 
in the absence of either large-scale damping or finite-size effects.
Focussing on the wave modes, we see that the energy transfer is 
downscale. Further, on equilibration the wave modes yield a well-defined smooth power-law. Given our modest
resolution, we see in Fig.\ (\ref{fig2}) that the 
wave-mode spectrum is consistent with a power law 
between $k^{-1}$ to $k^{-5/3}$ for $k_f < k < k_d$, where $k_f, k_d$ are 
the forcing and dissipation
wavenumbers. 
These results are consistent with previous forced and decaying simulations with 
$Bu \sim 1$ \cite{Bartello},\cite{SS-gafd},\cite{Kitamura}. 
Also, note that the results appear to have converged, i.e.\ we do not see any dependence of the scaling
and behavior of the modes for different numerical resolutions. Fig.\ (\ref{fig3}) shows the 
wave, vortical and total energy spectra for the two aforementioned resolutions.  \\

\begin{figure}
\centering
\includegraphics[width=10cm,height=10cm]{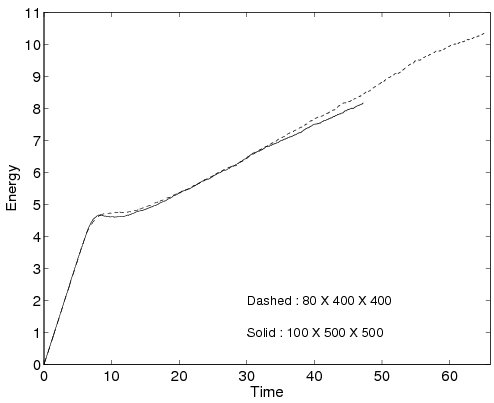}
\caption{\label{fig4} The total energy vs. time for the GGG model. The
simulations are at resolutions of $80 \times 400 \times 400$ and $100 \times 500 \times 500$ in an aspect
ratio of $1/5$ and $Ro=Fr=0.05$. Quite 
clearly, the system is not near equilibration; some of the runs were carried out significantly longer
times and yet showed no sign of settling down. The spectra (not shown) are also evolving in time and
do not show smooth power-law scaling.}
\end{figure}

It is well-known that the 2D stratified problem (which only supports wave modes), on large-scale 
forcing, yields a clear forward transfer of energy \cite{BS},\cite{SS}. Further, the kinetic 
equations derived from considering resonant interactions of wave modes support power-law 
solutions \cite{L1},\cite{L2}. Keeping in mind that the GGG model can be looked upon as a 3D extension of the 
2D stratified 
problem, these two pieces of information lead to the possibility that the GGG model will in fact 
support a forward
transfer of energy and yield an equilibrated power-law spectrum.  
Forcing the GGG model with method 2, we show
the total energy (equal to the wave-mode energy) vs.\ time in Fig.\ (\ref{fig4}).
Quite clearly, the energy does not equilibrate. 
On comparing the spectra (not shown), none of the runs showed a well-defined power-law. In essence, 
there is a forward transfer of energy in the GGG model, but it is highly inefficient compared to the full 
system forced by method 1. This inefficiency manifests itself in the lack of a systematic forward
transfer and the resulting lack of an (even approximate) inertial range 
as is observed in the full 
system.  For the given set of governing parameters, we speculate 
that the GGG model is probably under-resolved at a 
resolution when the full rotating Boussinesq system forced by method 1 yields consistent results across a range of resolutions. \\

\begin{figure}
\centering
\includegraphics[width=10cm,height=10cm]{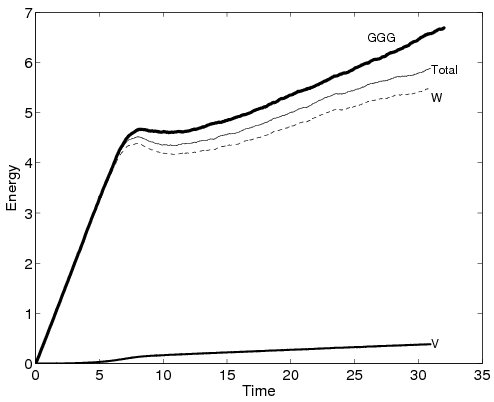}
\caption{\label{figaddon} The total energy vs. time for the full system forced by method 2. The resolution is 
$100 \times 500 \times 500$ in an aspect ratio of $1/5$ and $Ro=Fr=0.05$. The system does not equilibriate as it did when 
forced by method 1. However, due to the presence of the interactions among the vortical modes it does a better job at moving 
energy to small scales than the identically forced GGG model does. It is possible that both systems are under-resolved in this 
parameter regime.}
\end{figure}

It is interesting to observe how the presence of the interactions involving vortical modes compares to the results obtained from 
the GGG model when the full system is forced identically to the GGG model by method 2. Like the energy in the GGG model, the 
energy in the full system does not equilibriate Fig.\ (\ref{figaddon}). However, the full system does appear do a better job of 
moving the energy to small scales. After an initial growth of energy in the vortical modes ($\approx t=5$), the energy in the 
full system grows at a slower rate than in the GGG system. Like the GGG system, the full system now produces spectra (not shown) 
that are not well defined power-laws. In fact, the picture that emerges is that there is some transfer from the wave to 
vortical modes that results in 
the activation of $(w,v,w)$ 
interactions which help the full system move energy toward small scales. 
But, much like the GGG model the $(w,w,w)$ interactions are incapable of a robust forward transfer --- possibily due to their 
being under-resolved in
this set of parameters. \\

\begin{figure}
\centering
\includegraphics[width=8cm,height=8cm]{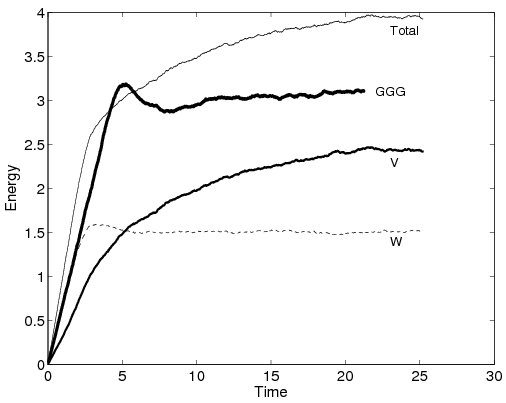}
\includegraphics[width=8cm,height=8cm]{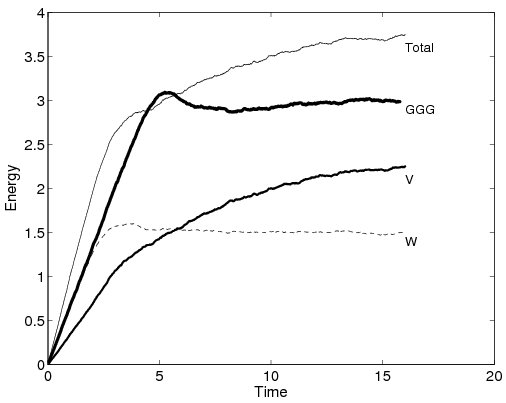}
\caption{\label{fig5} Left : $144\times 432\times 432$. Right : $162\times 486\times 486$. The total, vortical and wave-mode 
energy vs. time for the full model 
and the GGG model in an aspect ratio of $1/3$ with $Ro=Fr=0.1$. The behavior of the full model is similar to
Fig.\ (\ref{fig1}).  While, in contrast to Fig.\ (\ref{fig4}), now 
the GGG energy does level-off. Notice the agreement for both the full system and GGG between the two resolutions.}
\end{figure}

\begin{figure}
\centering
\includegraphics[width=8cm,height=8cm]{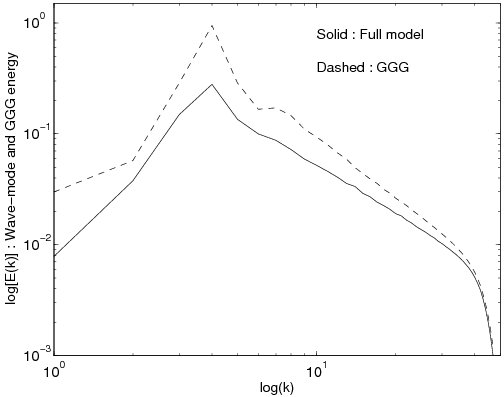}
\includegraphics[width=8cm,height=8cm]{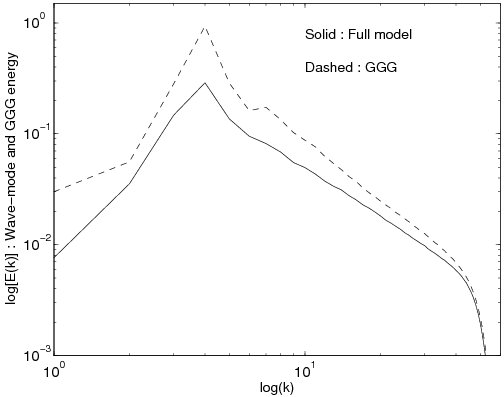}
\caption{\label{fig6} Left : $144\times 432\times 432$ ($\approx t= 17.5 s$). Right : $162\times 486\times 486$ 
($\approx t= 15.5 s$).  Wave-mode energy spectra 
for full and GGG models in an aspect
ratio $1/3$ with $Ro=Fr=0.1$. These are spectra from a single time snapshot of the models in the 
quasi-statistically steady range. Quite clearly, both may be interpreted as fairly clear power-laws with the GGG model spectrum 
being considerably steeper than the full system's wave-mode spectrum which scales between $-1$ and $-5/3$. }
\end{figure}

The notion that the GGG model supports a forward transfer of energy, albeit an inefficient one, but 
does not yield a steady state, is somewhat unsatisfactory. There is nothing intrinsically 
wrong with such a state of affairs, 
but given the possibility of their being under-resolved in the previous simulation, to gain more confidence in the GGG model 
we repeated the aforementioned experiment at higher 
$Fr,Ro$ (i.e.\ under weaker rotation and milder stratification) and in a less skewed domain. 
Specifically, setting $H/L=1/3$ and $Ro=Fr=0.1$, we performed simulations at resolutions of
$144 \times 432 \times 432$ and $162\times 486\times 486$. Note that the aspect ratio 1/3 case allows us to use a higher 
numerical resolution as compared to the aspect ratio 1/5 case. Fig.\ (\ref{fig5}) shows the energy in time for the GGG model 
forced by method 2 and the full model forced by method 1.  
In contrast to Fig.\ (\ref{fig4}), now the GGG energy levels-off fairly rapidly. Further, as is shown in Fig.\ (\ref{fig6}), 
the spectral distribution 
of the GGG energy is in the form of a smooth power-law.  It is interesting to note, from Fig.\ (\ref{fig6}), 
that the GGG spectrum is steeper than the full system's wave-mode spectrum.  In fact, this steepness
is qualitatively consistent with the resonant kinetic equation solutions \cite{L1},\cite{L2}. \\

\begin{figure}
\centering
\includegraphics[width=8cm,height=8cm]{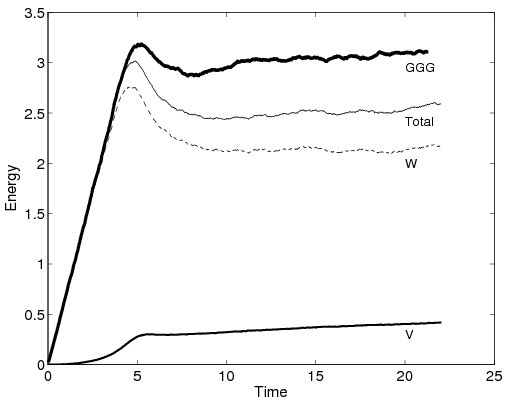}
\includegraphics[width=8cm,height=8cm]{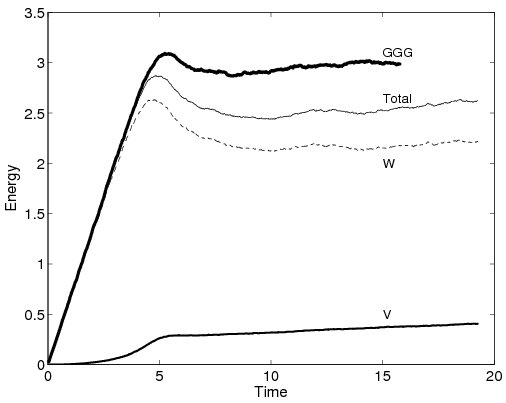}
\caption{\label{fig7} Left : $144\times 432\times 432$. Right : $162\times 486\times 486$. The GGG and full system wave, 
vortical and total energy in an aspect ratio $1/3$ and $Ro=Fr=0.1$ setting in which identical forcing by the IG modes was used. 
Quite clearly, both systems attain an energetically steady state. Notice the agreement for both the full system and GGG 
between the two resolutions.}
\end{figure}

\begin{figure}
\centering
\includegraphics[width=8cm,height=8cm]{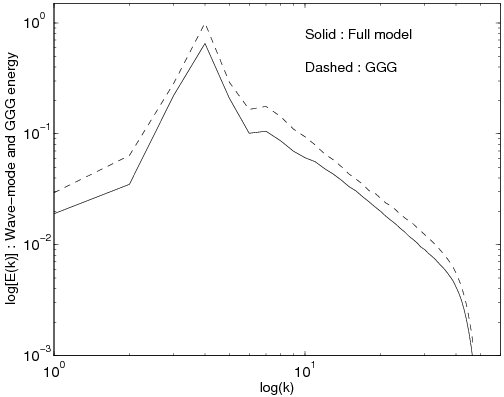}
\includegraphics[width=8cm,height=8cm]{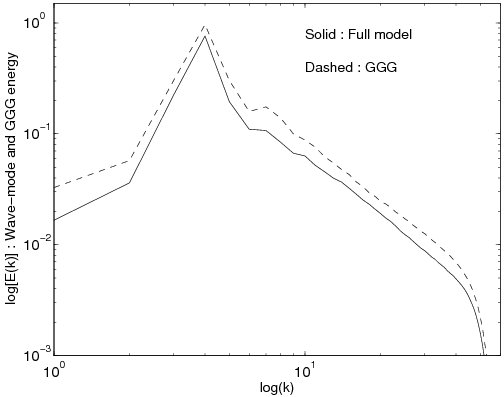}
\caption{\label{fig8} Left : $144\times 432\times 432$ ($\approx t= 18.65 s$). Right : $162\times 486\times 486$ 
($\approx t= 17.7 s$). The GGG and full system wave-mode energy spectrum in an aspect 
ratio $1/3$ and $Ro=Fr=0.1$ setting in which identical forcing by the IG modes was used. These are spectra from a single time 
snapshot of the models. Quite clearly, both may be interpreted as fairly consistent power-laws.} 
\end{figure}

As the GGG model equilibrates in these milder conditions, it is of interest to 
explore the response of the full rotating Boussinesq system forced the same as the GGG model by method 2. 
Note that when the full system is forced by method 1 there was a natural flow of energy into the vortical mode that allowed 
the vortical mode to mediate energy transfer between 
two wave modes by means of wave-vortical-wave interactions. Given that 
wave to vortical transfers are weak (i.e.\ only due to near-resonances), 
forcing only the wave-modes allows us to see if 
the $(w,w,w)$ interactions in the full system behave in accord 
with the GGG model. This is precisely the case, and along with the attainment of a steady state
(Fig. (\ref{fig7})), both simulations yield almost identical power-law 
wave-mode spectra for $Ro=Fr=0.1$ with $H/L=1/3$ (see Fig (\ref{fig8})). \\

In essence, this confirms our speculation that that the wave-mode interactions were under-resolved in our numerical simulations of
rapid rotation and strong stratification
(specifically, $Ro,Fr \le 0.05$) in the more skewed domain ($H/L < 1/5$). Quite naturally, the wave-mode
energy distribution in the
full rotating Boussinesq system is then primarily controlled by the $(w,w,v)$ interactions \cite{WB2}.
As also noted by Waite and Bartello \cite{WB2}, our computations suggest the possibility
that present-day numerical models may be deficient in their representation of
$(w,w,w)$ interactions when trying to work in geophysically relevant parameter regimes.
Further, in milder conditions (i.e.\ $Ro,Fr=0.1$ and $H/L=1/3$), the GGG model resulted in a
steady state with a well-defined power-law energy distribution. This leads us to the possibility that,
on proper resolution, the $(w,w,w)$ interactions may play a role in the wave-mode energy transfer and
distribution in realistic parameter regimes. Future work with better computational resources will inquire 
into this possibility as well as 
universality in the
GGG spectral scaling. \\

\section{Conclusions}

In rotating and stratified flows, the forward transfer and distribution of 
energy among wave modes is controlled primarily by wave-vortical-wave 
$(w,w,v)$ and wave-wave-wave $(w,w,w)$
interactions. However, the relative importance of these two classes of interactions 
is somewhat unclear. On one hand, prior analytical \cite{Babin-rev} and 
numerical \cite{Bartello},\cite{SS-gafd} work suggests that the wave-vortical-wave class of interactions is of
primary importance --- when $1/2 \le f/N \le 2$, this is immediately evident  \cite{WB2}.  
On the other hand, 
especially outside the range $1/2 \le f/N \le 2$ (as is true in most geophysical scenarios), the kinetic
equation approach considering only resonant IG wave-mode interactions yields solutions that 
are consistent with certain observations \cite{L1},\cite{L2}. 
In order to inquire into this somewhat dichotomous situation, we constructed a reduced model
consisting only of wave-mode interactions (resonant and non-resonant).  
This GGG model is given by a PDE that conserves energy in the inviscid, unforced case. 
We studied the GGG model in
an idealized periodic setting, though its applicability is not limited 
to such idealized domains. \\

We compared the GGG model to the full system in a geophysically relevant parameter regime. Specifically, with a skewed aspect ratio $H/L< 1$, 
we chose $f,N$ such that $Ro=Fr \Rightarrow Bu=1$. Further, 
the flow was rapidly rotating and strongly stratified, i.e.\ $Fr=Ro \ll 1$. 
In the full system, as anticipated we observed an inverse transfer of energy among the 
vortical modes accompanying by a power-law that was consistent with a $k^{-3}$ scaling 
for $k_f < k < k_d$. The wave-mode energy was transferred to small scales, and these 
modes equilibrated quite rapidly.   In accord with prior work \cite{Bartello},\cite{SS-gafd},\cite{Kitamura}, the wave-mode energy distribution was 
consistent with a power law with scaling between a $k^{-1}$ to $k^{-5/3}$ form for $k_f < k < k_d$. \\ 

For the GGG model, we also anticipated a statistically steady wave-mode energy
spectrum with power-law scaling. This expectation follows from previous results 
establishing that (i) power-law solutions are supported by the kinetic equations governing resonant wave-mode interactions, and (ii) the GGG model is a 3D extension of the 2D stratified problem 
(known to support a robust forward transfer of energy).   A clear power law was 
indeed observed for "mild" conditions with $H/L=1/3$ and $Fr =Ro \approx 0.1$.  
 Consistent with the kinetic equations for resonant
wave interactions, the GGG energy spectra were observed to be steeper than the
wave-mode energy spectra associated with the full equations when all modes
were forced with equal weight. When the forcing was restricted to excite
only wave modes, then the GGG and full systems yielded wave-mode
spectra with essentially identical power-law scaling.
The results indicate that $(w,w,w)$ interactions can play a significant
role in the transfer of energy from forced scales to smaller scales, especially
when the wave modes are preferentially excited.
However, the energy did not equilibrate for the GGG flow in the
smaller aspect ratio $H/L = 1/5$ with stronger rotation and stratification ($Fr=Ro = 0.05$), suggesting
an in-efficient forward transfer of energy.  Furthermore, the full system also did
not equilibrate for these parameter values when only the wave modes
were excited at large scales.  These results serve as a caution that 
improper resolution of wave-mode interactions may be a significant issue in present-day
numerical models that attempt to work in geophysically relevant parameter regimes,
consistent with \cite{WB2}. Finally, there are a number of issues that raised herein that we feel merit further examination. 
For example, is there universality in the wave-mode spectrum from the GGG model (with respect to decreasing $Fr=Ro$ in a 
fixed aspect ratio). Similarly, with better numerical resources, an important issue to be addressed is the difference in 
scaling of the wave-mode energy from full model simulations (with forcing of type 1) and that from the GGG model. 
This gets to the heart of the matter with regard to energy distribution resulting from $(w,v,w)$ and $(w,w,w)$ interactions
respectively --- indeed, the intriguing possibility that the scaling from these two scenarios may converge 
with progressively stronger rotation and
stratification provides adequate motivation for such an exploration.

\vskip 1 truecm
{\it Acknowledgements : }
Financial support
was provided by NSF
CMG 0529596 and
the DOE Multiscale Mathematics program (DE-FG02-05ER25703). \\

\section{Appendix}

Here we present the GGG model by utilizing the eigenfunction decomposition suggested in (\ref{represent}).  For the general case, i.e.\ $k_x,k_y,k_z \neq 0$, the eigenfunctions are  

\begin{eqnarray}
\phi^+=\frac{1}{\sqrt{2}\sigma_kk}\left(\begin{array}{c}\frac{k_z}{k_h}(\sigma_kk_x+{\rm i}k_yf)\\ 
\frac{k_z}{k_h}(\sigma_kk_y- {\rm i}k_xf)\\ -\sigma_kk_h\\- {\rm i}Nk_h\end{array}\right),
\quad\phi^-={\phi^+}^*,
\quad\phi^0=\frac{1}{\sigma_kk}\left(\begin{array}{c}Nk_y \\-Nk_x \\0 \\ fk_z\end{array}\right)
\label{4}
\end{eqnarray}
here ${(\cdot)}^*$ denotes complex-conjugate.  The special case, $k_h=0, k_z \neq 0$ (the 
VSHF mode) is treated separately.  Here $\sigma^0=0, ~\sigma^{\pm}(\BO{k})=\pm f$ and  

\begin{eqnarray}
\phi^+=\left(\begin{array}{c}\frac{{\rm i}}{\sqrt{2}}\\ \frac{1}{\sqrt{2}}\\ 0\\0 \end{array}\right),
\quad\phi^-={\phi^+}^*,
\quad\phi^0=\left(\begin{array}{c}0 \\ 0 \\0 \\\text{sgn}(fk_z)\end{array}\right) 
\label{5}
\end{eqnarray}
or,
\begin{eqnarray}
\phi^+=\frac{1}{2}\left(\begin{array}{c}1+{\rm i}\\1-{\rm i}\\ 0\\0 \end{array}\right),
\quad\phi^-={\phi^+}^*,
\quad\phi^0=\left(\begin{array}{c}0 \\ 0 \\0 \\\text{sgn}(fk_z)\end{array}\right)
\label{6}
\end{eqnarray}
Substituting in (\ref{represent}) and using the orthogonality of the modes, with $\psi$ as the 
streamfunction, we have

\begin{itemize}
\item For $k_x,k_y,k_z \neq 0$ : 
\begin{eqnarray}
a_k^0=\frac{{\rm i}N}{\sigma_kk} (-k_h^2\psi- {\rm i}\frac{f}{N}k_z\theta) \nonumber \\
a_k^+=\frac{1}{\sqrt{2}\sigma_kk}(-\frac{\sigma_k w_k k^2}{k_h}-fk_z k_h\psi+ {\rm i}Nk_h\theta) \nonumber \\
a_k^-=\frac{1}{\sqrt{2}\sigma_kk}(-\frac{\sigma_k w_k k^2}{k_h}+fk_z k_h\psi- {\rm i}Nk_h\theta) \nonumber \\
a_k^++a_k^-=-\frac{\sqrt{2}kw_k}{k_h} \nonumber \\
a_k^+-a_k^-=\frac{\sqrt{2}{\rm i}fk_h}{\sigma_kk}(\frac{N}{f}\theta+{\rm i}k_z\psi)
\label{7}
\end{eqnarray}
\item For $k_h=0, k_z \neq 0$ :
\begin{eqnarray}
a_{k_z}^0=\overline{\theta(z)}\text{sgn}(fk_z)\quad a_{k_z}^+=\frac{-{\rm i}\overline{u}+\overline{v}}{\sqrt{2}}
\quad a_{k_z}^-=\frac{{\rm i}\overline{u}+\overline{v}}{\sqrt{2}} \nonumber \\
(a^+_{k_z}+a^-_{k_z})=\sqrt{2}\overline{v(z)}\qquad (a^+_{k_z}-a^-_{k_z})=-{\rm i}\sqrt{2}\overline{u(z)}
\label{8}
\end{eqnarray}
where $\overline{(\cdot)}$ denotes a horizontal average.
\end{itemize}
Using (\ref{4}) and (\ref{7}), the GGG model (retaining only IG mode interactions) for the 
$k_x,k_y,k_z \neq 0$ case is

\begin{eqnarray}
\PD{(a_\BO{k}^++a_\BO{k}^-)}{t}&+{\rm i}\sigma_\BO{k}(a^+_\BO{k}-a^-_\BO{k})=\sum_{\BO{k}=\BO{p}+\BO{q}}\frac{1}{\sqrt{2}\sigma_kk}\Bigg\{(\frac{a_\BO{p}^++a_\BO{p}^-}{\sigma_pp}) 
\left(\frac{a_\BO{q}^++a_\BO{q}^-}{\sigma_qq}\right)\Big\{\frac{{\rm i}\sigma_p\sigma_q\sigma_kp_h^2q_h^2p_z^3}{p_hq_hk_h} \nonumber \\
&+\frac{{\rm i}\sigma_p\sigma_q\sigma_kp_h^2q_h^2p_zq_z^2}{p_hq_hk_h}+\frac{{\rm i}\sigma_p\sigma_q\sigma_k(\BO{p}_h\cdot\BO{q}_h)q_h^2p_z^3}{p_hq_hk_h}-\frac{{\rm i}\sigma_p\sigma_q\sigma_k(\BO{p}_h\cdot\BO{q}_h)q_h^2p_zq_z^2}{p_hq_hk_h})
\nonumber \\
&-2\frac{{\rm i}\sigma_p\sigma_q\sigma_k(\BO{p}_h\cdot\BO{q})^2p_zq_z^2}{p_hq_hk_h}+{\rm i}\sigma_p\sigma_q\sigma_kp_hq_hk_hp_z-{\rm i}\sigma_p\sigma_q\sigma_k(\BO{p}_h\cdot\BO{q}_h)\frac{q_h}{p_h}p_z\}\Big\} \nonumber \\
&+\left(\frac{a_\BO{p}^++a_\BO{p}^-}{\sigma_pp}\right)\left(\frac{a_\BO{q}^+-a_\BO{q}^-}{\sigma_qq}\right)\Big\{2\frac{f\sigma_k\sigma_p(\BO{p}\times\BO{q}\cdot\hat{z})(\BO{p}_h\cdot\BO{q}_h)(p_z^2q_z+p_zq_z^2)}{p_hq_hk_h}
\nonumber \\
&-\frac{f\sigma_p\sigma_kq_z^3(\BO{p}\times\BO{q}\cdot\hat{z})p_h^2}{p_hq_hk_h}+\frac{f\sigma_k\sigma_pp_z^2q_z(\BO{p}\times\BO{q}\cdot\hat{z})p_h^2}{p_hq_hk_h}+f\sigma_k\sigma_pk_h(\BO{p}\times\BO{q}\cdot\hat{z})\frac{p_h}{q_h}q_z\Big\} \nonumber \\
&+\left(\frac{a_\BO{p}^+-a_\BO{p}^-}{\sigma_pp}\right)\left(\frac{a_\BO{q}^+-a_\BO{q}^-}{\sigma_qq}\right)\Big\{-2\frac{{\rm i}f^2\sigma_k(\BO{p}\times\BO{q}\cdot\hat{z})^2p_z^2q_z}{p_hq_hk_h}\Big\}\Bigg\}
\label{9}
\end{eqnarray}
and 

\begin{eqnarray}
\PD{(a_\BO{k}^+-a_\BO{k}^-)}{t}&+{\rm i}\sigma_\BO{k}(a^+_\BO{k}+a^-_\BO{k})=\sum_{\BO{k}=\BO{p}+\BO{q}}\frac{1}{2\sqrt{2}\sigma_kk}\Bigg\{\left(\frac{a_\BO{p}^++a_\BO{p}^-}{\sigma_pp}\right)\left(\frac{a_\BO{q}^++a_\BO{q}^-}{\sigma_qq}\right)\Big\{\frac{
f\sigma_p\sigma_q(\BO{p}\times\BO{q}\cdot\hat{z})p_z^3q_h^2}{p_hq_hk_h} 
\nonumber \\
&+\frac{f\sigma_p\sigma_q(\BO{p}\times\BO{q}\cdot\hat{z})p_z^2q_zq_h^2}{p_hq_hk_h}\Big\}+\left(\frac{a_\BO{p}^++a_\BO{p}^-}{\sigma_pp}\right)\left(\frac{a_\BO{q}^+-a_\BO{q}^-}{\sigma_qq}\right)\Big\{-\frac{{\rm i}f^2\sigma_pp_z^2q_zp_h^2q_h^2}{p_hq_hk_h}-
\frac{{\rm i}f^2\sigma_pp_z^2q_z(\BO{p}_h\cdot\BO{q}_h)q_h^2}{p_hq_hk_h} \nonumber \\
&-\frac{{\rm i}f^2\sigma_pp_zq_z^2(\BO{p}_h\cdot\BO{q}_h)q_h^2}{p_hq_hk_h}+\frac{{\rm i}f^2\sigma_pq_z^3p_h^2q_h^2}{p_hq_hk_h}+\frac{{\rm i}f^2\sigma_pp_zq_z^2(\BO{p}_h\cdot\BO{q}_h)p_h^2}{p_hq_hk_h}+\frac{{\rm i}f^2\sigma_pq_z^3(\BO{p}_h\cdot\BO{q}_h)p_h
^2}{p_hq_hk_h} \nonumber \\
&-{\rm i}N^2k_h(\BO{p}_h\cdot\BO{q}_h)\frac{q_h}{p_h}p_z\sigma_p+{\rm i}N^2p_hq_hk_hq_z\sigma_p\Big\}+\left(\frac{a_\BO{p}^+-a_\BO{p}^-}{\sigma_pp}\right)\left(\frac{a_\BO{q}^+-a_\BO{q}^-}{\sigma_qq}\right)\Big\{ 
\nonumber \\
&+\frac{f^3(\BO{p}\times\BO{q}\cdot\hat{z})p_z^2q_zp_h^2}{p_hq_hk_k}+\frac{f^3(\BO{p}\times\BO{q}\cdot\hat{z})p_zq_z^2p_h^2}{p_hq_hk_k}+fN^2k_h(\BO{p}\times\BO{q}\cdot\hat{z})q_z\frac{p_h}{q_h}\Big\}\Bigg\}
\label{10}
\end{eqnarray}
The VSHF contributions to the above are 

\begin{eqnarray}
\PD{(a_\BO{k}^++a_\BO{k}^-)}{t}&+{\rm i}\sigma_\BO{k}(a^+_\BO{k}-a^-_\BO{k})=\sum_{k_z=p_z+q_z,k_h=q_h}\frac{1}{\sqrt{2}k}\Big\{(a^+_{p_z}+a^-_{p_z})(\frac{a_\BO{q}^++a_\BO{q}^-}{q})({\rm i}q_yp_z^2-{\rm i}q_yq_z^2-{\rm i}q_yq_h^2) 
\nonumber \\
&+(a^+_{p_z}-a^-_{p_z})(\frac{a_\BO{q}^++a_\BO{q}^-}{q})(-q_xp_z^2+q_xq_z^2+q_xq_h^2)\Big\}
\label{11}
\end{eqnarray} 
and

\begin{eqnarray}
\PD{(a_\BO{k}^+-a_\BO{k}^-)}{t}&+{\rm i}\sigma_\BO{k}(a^+_\BO{k}+a^-_\BO{k})=\sum_{k_z=p_z+q_z,k_h=q_h}\frac{1}{\sqrt{2}\sigma_kk}\Big\{(a^+_{p_z}+a^-_{p_z})(\frac{a_\BO{q}^++a_\BO{q}^-}{q})(-fq_xp_z(p_z+q_z) 
\nonumber \\
&+(a^+_{p_z}-a^-_{p_z})(\frac{a_\BO{q}^++a_\BO{q}^-}{q})({\rm i}fq_yp_z(p_z+q_z))+(a^+_{p_z}+a^-_{p_z})(\frac{a_\BO{q}^+-a_\BO{q}^-}{\sigma_qq})(-{\rm i}f^2q_yq_z(p_z+q_z) \nonumber \\
&-{\rm i}N^2q_yq_h^2)+(a^+_{p_z}-a^-_{p_z})(\frac{a_\BO{q}^+-a_\BO{q}^-}{q\sigma_q})(f^2q_xq_z(p_z+q_z)+N^2q_yq_h^2)\Big\}
\label{12}
\end{eqnarray}
and finally, using (\ref{5}) and (\ref{9}), the evolution the VSHF modes themselves is given by 

\begin{eqnarray}
\PD{(a^+_{k_z}+a^-_{k_z})}{t}+{\rm i}\sigma_{k_z}(a^+_{k_z}-a^-_{k_z})&=-\sum_{k_z=p_z+q_z,\BO{p_h}}\frac{1}{2\sqrt{2}}\Bigg\{(\frac{a^+_p+a^-_p}{p})(\frac{a^+_{\BO{p_h},qz}+a^-_{\BO{p_h},qz}}{q})({\rm i}p_y(q_z^2-p_z^2) 
\nonumber \\
&+(\frac{a_p^+-a_p^-}{\sigma_pp})(\frac{a^+_{\BO{p_h},qz}+a^-_{\BO{p_h},qz}}{q})(-fp_xp_zq_z-fp_xp_z^2) 
\nonumber \\
&+(\frac{a_p^++a_p^-}{p})(\frac{a^+_{\BO{p_h},qz}-a^-_{\BO{p_h},qz}}{\sigma_qq})(fp_xp_zq_z+fp_xq_z^2)\Bigg\}
\label{13}
\end{eqnarray}
and

\begin{eqnarray}
\PD{(a^+_{k_z}-a^-_{k_z})}{t}+{\rm i}\sigma_{k_z}(a^+_{k_z}+a^-_{k_z})&=-\sum_{k_z=p_z+q_z,\BO{p_h}}\frac{1}{2\sqrt{2}}\Bigg\{(\frac{a^+_p+a^-_p}{p})(\frac{a^+_{\BO{p_h},qz}+a^-_{\BO{p_h},qz}}{q})p_x(q_z^2-p_z^2) \nonumber \\
&+(\frac{a_p^+-a_p^-}{\sigma_pp})(\frac{a^+_{\BO{p_h},qz}+a^-_{\BO{p_h},qz}}{q})(-{\rm i}p_yf(p_z^2+p_zq_z)) 
\nonumber \\
&+(\frac{a_p^++a_p^-}{p})(\frac{a^+_{\BO{p_h},qz}-a^-_{\BO{p_h},qz}}{\sigma_qq}){\rm i}p_yf(q_z^2+p_zq_z)\Bigg\}.
\label{14}
\end{eqnarray}
It can be 
verified that inverse transforming these equations leads to exactly the set specified in Section IIB. \\

\end{document}